\documentstyle[aps,aipbook,floats]{revtex}
\newcommand{\be}{\begin{equation}}
\newcommand{\ee}{\end{equation}}
\begin{document}
\title{Precessing Gamma Jets in extended and evaporating Galactic Halo as 
source of GRB}

\author{Daniele Fargion$^*$ , Andrea Salis}
\address{Physics Dept. Rome University "La Sapienza"\\
P.le A.Moro 00187\\ 
$^*$INFN sez. Roma1 Univ."La Sapienza", Rome}

\maketitle

\begin{abstract}
The Precessing Gamma Jets (GJ) in binary systems located in extended or 
evaporating galactic halo should be the source of GRB. The GJ are born by 
Inverse Compton Scattering (ICS) of thermal photons (optical, infrared,...) 
onto (power law) electron jets (from GeV energies and above) produced by 
spinning pulsars or black holes. The thermal photons are emitted by the binary 
companion (or by their nearby accreting disk). The collimated GJ beam is 
trembling with the characteristic pulsar millisecond period and it is bent by 
the companion magnetic field interactions, as a lighthouse, in a nearly 
conical shape within the characteristic Keplerian period; an additional 
nutation due to the asymmetric inertial momentum may lead, in general, to 
aperiodic behaviour of GRB signals. SGRs are GRBs seen at the periphery of the 
hard energy  GJ beam core. The original birth locations of GJ (SNRs, planetary 
nebulae, globular clusters,...) are smeared out by the high 
escape velocity of the system; the Neutron Star (NS) high velocity is possibly 
due to the asymmetric jet precession , and consequent "rowing" 
acceleration, related to the eccentricity of the binary system. The GJ power 
is, for realistic parameters, comparable to the one needed for GRB in extended 
or evaporating  galactic halo. Their detailed spectra and time evolution fit 
the observed data. The expected GRB source number (within present BATSE 
sensitivity) is tens of thousands, compatible with the allowed 
presence of $10-20\%$ GRB repeaters. 
\end{abstract}

\section*{}

The Eddington luminosity and the opacity in GRBs are the cornerstones 
in understanding, constraining and finally building in a chain of arguments
 any realistic GRB model. Spherically symmetric GRBs, either in extended 
galactic halo or at the cosmological edges, overcome Eddington luminosity 
(because of the millisecond times and the hundred Kms spaces of GRB structure).
 The consequent explosive fireball models imply a copious electron pairs
 production (by $\gamma-\gamma$ scattering), huge opacity and a consequent
 photon thermalization (via electron pairs) which is in disagreement with 
the observed non thermal GRB spectra. 
This over Eddington luminosity (and opacity) problem is present also for the 
SGRs as isotropic sources. In this framework it is not possible to solve the 
problem by splitting or hiding the GRBs nature from the SGRs one; on the
contrary this "discriminatory" attitude just doubles and deepens the puzzle. 
Moreover the unique gamma event, the 5/3/79 GRB [1] its earliest hard spectra 
nature implies and calls for a strong link between SGRs and GRBs. Consequently 
the SGR locations suggest a "local" (extended galactic halo, local group) 
site for GRBs. Therefore the main steps toward the GRB model are: 1) the 
fireball (isotropic) GRB sources (galactic or 
cosmological) are thermal sources; we need a non thermal one and we consider 
a non equilibrium gamma (anisotropic) {\it beamed source}. 2) The high p-p 
scattering as gamma source leads to characteristic pion mass energy spectra 
absent in GRB spectral breaks, lines or other features. Moreover these 
scattering exhibits a poor beaming efficiency ($\theta_{\gamma}\simeq
\theta_{\pi^o}\simeq\sqrt{\frac{m_p}{E_p}}$) and lower flux amplification 
compared, for example, to the Inverse Compton Scattering (ICS) by electrons at 
same energy ($\theta_{\gamma}\simeq\theta_e\simeq\frac{m_e}{E_e}$). 3) 
Electron-positron pair annihilation (and 
their generation) should be traced by characteristic gamma lines (even smeared 
by Doppler shift) nearly absent in most known GRB spectra. 4) Synchrotron 
radiation at MeV leads to spectra in disagreement with data and it calls for 
unbelievable huge magnetic fields and/or cosmic electron energies as well as
long collimation lenght.
 5) Bremsstrahlung radiation as above is also in disagreement with 
observational GRB data (no beaming and consequent thermalization).
 6) We were forced to consider the best physical process for beaming GRB [2]: 
the ICS of ultrarelativistic electrons onto thermal photons (we neglected 
the ICS contribute by protons because of their severe suppression factor in 
the corresponding Thomson cross section: ($(\frac{m_p}{m_e})^2$). 7) Because 
of the primary cosmic rays electron spectrum near the Earth we considered the 
energy composition of the electrons near the maximum at GeV as the electron 
beam characteristic energy in jets. 8) In this way we assumed that the same 
GeV electron jets are responsible for a large fraction of primary cosmic rays. 
The jet number is therefore bounded by the total cosmic rays electrons flux 
data. These anisotropic beams of c.r. electrons are now observed isotropically 
because of the presence of randomizing and smoothing interstellar magnetic 
fields. 9) The absence of any detectable anisotropy for c.r. electrons at GeV 
also implies a limited ($<$ yrs or tens of yrs) collimated lenght of the 
electron jets. 10) Relativistic kinematics of ICS leads, for any large c.r. 
electrons $\gamma$ Lorentz factor, to a corresponding small beam angle 
$\theta_e$ for the ICS gamma rays favouring the existence of an associate 
collimated gamma beam or GJ. The characteristic Lorentz factor, the 
consequent gamma beam collimation and the average gamma energies are
$$
\gamma_e=2\cdot 10^3\Big(\frac{E_e}{GeV}\Big);~~~~ \theta_e=5\cdot 10^{-4}\Big(
\frac{E_e}{GeV}\Big)^{-1}; 
$$
\be
\overline{h\nu_{\gamma}}=1.38\Big(\frac{\gamma_e}{2\cdot 
10^3}\Big)^2\Big(\frac{T}{6000~K}\Big) MeV
\ee
in rough agreement with the needed energies in GRBs. Here $E_e$ is the 
electron energy in the jet while T is the thermal temperature of the 
nearby companion star. 11) The existence of collimated blazar-like jets of 
particles is not new in astrophysics and it is known since long time in 
extragalactic high energy astrophysics (superluminal objects, quasars, etc.); 
their ICS, for instance onto cosmological BBR or self ICS on synchrotron 
radiation, may be the source of variable highest 
gamma rays as MRK 421. However the same extragalactic AGNs or quasars are not 
suitable GRB candidates because of their too large masses and because of the 
associated too long Schwartzchild time scales $\frac{r_s}{c}\sim 10^3\Big(
\frac{M}{10^8 M_\odot}\Big) s$, {\it i.e.} million times longer respect to 
the characteristic millisecond GRB time structures. 12) This millisecond 
trembling (or hundredth, tenth of second) of GRBs calls, let say better 
$\underline{cries}$, desperately for the 
pulsar-like (NS or BH) nature of the GRB sources. Indeed the (nearly) absence 
of submillisecond structure (if not a misleading coincidence of a perverse 
Nature...) is in full resonance with the lower known bounds on pulsar period 
at half millisecond. 13) Therefore we are forced [2] to assume as candidate 
sources of GRB compact fast spinning NS (or BH) which eject electrons (as well 
protons) at GeV energies (NSJ) and which also eject, by ICS, an associated 
gamma jet. The simplest hypothesis, ICS onto 2.73 K BBR, leads to a quite 
unefficient electron jet $\rightarrow$ GJ energy conversion in disagreement 
with needed intensities in GRBs. 14) Therefore, in order to enhance and 
amplify the ICS efficiency to the needed ones, we are driven to assume a 
nearby "lamp", {\it i.e.} a nearby copious stellar photon source as a binary 
companion. 15) The binary companion plays also a key role in deflecting and in 
precessing and possibly powering the jet leading to dynamical "blazing" or 
"lighthouse" beaming which explains the transient "explosive" nature of the 
GRBs. 16) Therefore the ICS onto thermal photons gives life to a $\underline{
collinear}$ GJ along with the NSJ. We remind that these NSJ or GJ are in 
general symmetric (up-down) jets. Moreover, as we show elsewhere [3], the 
internal angular structure of the GJ (by relativistic kinematics) allows the 
understanding of the time-energy evolution (soft-hard-soft) of GRB spectra 
during the sweeping of the precessing GJ. 17) The compact object, source of 
the jet, NSJ, of a mass $M_c$, in the Keplerian system, with a solar-like 
companion of mass $M_s$, precedes because of its dipolar magnetic interaction 
with the companion field. 18) The symmetric (up-down) rotating GJ sprays in a 
conical shape as a lighthouse in the dark spaces. It becomes observable (as 
just a laser light in a night club) either if a large gas or dust volume (or a 
shell screen) diffuses or reflects it, 
or if the observer enters inside the beam core. Recent astrophysical 
candidates in such configurations have been revealed: the last observed twin 
diffused cones in Egg Nebula (CRL 2688), the twin rings in SN1987A and in 
Hourglass Nebula projected on their red giant relic shells. 19) The Keplerian 
angular velocity $\omega_b$ must be derived by the characteristic $\theta_e$ 
beam angle and by the GRB duration lenght $\Delta\tau_b$; therefore it also 
underlines a characteristic consequent binary distance $r_b$:
\be
\omega_b\approx\frac{\theta_e}{\Delta\tau_b}\approx2\cdot 10^{-4} \Big(
\frac{\gamma_e}{2\cdot 10^3}~\frac{\Delta\tau_b}{2 s}\Big)^{-1} s^{-1}
\ee
\be
r_b\approx 2.5 c \Big[\Big(\frac{M_c}{M_\odot}\Big)+\Big(
\frac{M_s}{M_\odot}\Big)\Big]^{1/3}\cdot\Big(\frac{\gamma_e}{2\cdot 
10^3}~\frac{\Delta\tau_b}{2 s}\Big)^{2/3} s~~~.
\ee
20) These distances are characteristic of the Roche capture lobes of stellar 
sizes; so one must expect that these configurations lead to a feeding 
processes (accretion disk) and to a "strip tease" of the companion as well as 
to a merging of the companion onto the NSJ. 21) During that merging ($\Delta 
\tau\sim 2 s$, $r_b\sim 2 s$) an obscured (opaque) configuration would result 
and GRBs in those parameter window might be suppressed. 21) Indeed the 
characteristic GRB duration $\Delta\tau_b\sim 2 s$ exhibits an absence or a 
lack of sources: the transient opacity at that stages may be the cause of it. 
22) The consequent new configurations (smaller $\Delta\tau_b<2 s$ durations) 
which may arise in more bound systems are associated to smaller size white 
dwarf (or even NS) companions; the last stages may be compact and faster relic 
accreting disk (or ring) whose thermal photons are interacting (by ICS) onto 
the NSJ. 23) These more stable configurations might be associated to type II 
GRBs whose shorter durations, harder spectra are related to the higher Lorentz 
factors and the consequent more beamed jets. 24) The last oldest stable GJ may 
lead to periodic configurations and they may be identified with known gamma 
ray repeaters or even to gamma ray pulsars [2]. Recent evidence for such a 
strong link between SGRs and 
gamma ray (or hard X ray) pulsars are offered by the last discover [4] and the 
step by step (soft gamma ray burst $\rightarrow$ gamma pulsar) understanding 
of identified GROJ1744-28 pulsar, near the Galactic Center. Its 
luminosity nearly above the Eddington one implies a beaming (even if not as 
collimated and energetic as harder gamma ray pulsars or SGRs or GRBs). Indeed 
this last X pulsar is shining at peak luminosity ten times above Eddington 
critical luminosity for a solar mass $L_{cr}\sim 1.3\cdot 10^{38}\Big
(\frac{M}{M_\odot}\Big) erg~s^{-1}$; moreover its spinning down power 
$\dot{E}_s=-4\pi^2 I\frac{\dot{P}}{P^3}\sim 2.8\cdot 10^{36} erg~s^{-1}$ is 
nearly three order of magnitude $\underline{below}$ the observed peak or 
average luminosity $(3.5~or~1)\cdot 10^{39} erg~s^{-1}$, calling for an 
"exceptional" (the largest ever observed) conversion efficiency from 
rotational energy into X burst. This over luminosity may be explained, within 
our model of a binary GJ with stellar companion, assuming an emission jet 
nature of the pulsar whose beaming 
amplifies the nominal luminosity by a $\theta_b^{-2}$ factor. Moreover the 
orbital binary parameters just discovered for GROJ1744-28 (Roche lobe radious 
size $\sim 8 s\cdot c$) are well consistent with our model. 25) On 
the other extreme slow and wide binary systems are less powered (by accretion) 
and their jet energy is wider and it is therefore less observable (type I GRBs 
with longest burst duration). 26) As it will be shown in detail in the next 
article [3], because of the concentric cone energy distribution (soft-hard) of 
the beam by ICS, the SGRs might be due to the more probable 
observations of the peripheric (and softer) zones of those cone beams, while 
the most rare observation of the beam core (where the hardest component of the 
ICS spectra is hidden) leads to the rarest, hardest and powerful GRB event. 
Therefore too far away GRB sources cannot be seen at their "SGR stage" because 
of the BATSE sensitivity threshold. 27) For a quantitative evaluation of the 
model one easely derives [2] the NSJ electron power $\dot{E}_j$ conversion (by 
ICS) into the beamed GJ one $\dot{E}_{\gamma}$
\be
\frac{dE_{\gamma}}{dt}\approx 5\cdot 10^{41}\Big(\frac{\gamma_e}{2\cdot 
10^3}\Big)^3\Big(\frac{T}{6000 K}\Big)^4\Big(\frac{r_b}{2.5 s\cdot c}\Big)^{-
1}\Big(\frac{\dot{E}_j}{10^{38} erg s^{-1}}\Big)\Big(\frac{r_s}{R_\odot}\Big)
^2 erg s^{-1}
\ee
where $r_b$ is the binary distance of eq.3, T and $r_s$ are the companion star 
temperature and radius, $\gamma_e$ is the Lorentz factor of eq.1. 28) The 
assumed $\dot{E}_j$ power is a characteristic calibrating one for known 
galactic jets as Great Annihilator or SS433; one or even two order of 
magnitude are uncertain and they may be freely scaled on the $\dot{E}_j$ 
powers. Indeed these GJ power are characteristic of GRB in extended ($\sim 100 
Kpc$) or giant evaporating ($\sim 500 Kpc$) volumes (or even within local 
group volumes). These huge volumes satisfy the observed GRB isotropy and the 
"edge" inhomogeneity. 29) The merging with Andromeda halo must and might be 
observed in a near future by some kind of anisotropy in GRB distribution at 
lowest fluxes. More details on the GRB model, spectra, applications and 
observational evidences are shown in ref.[3]; our model gives spectra able to 
successfully fit observed GRB ones and their "thermal evolution" 
contrary to quasi thermal fireball spectrum. Finally the relic c.r. 
electrons by NSJ are possibly related to the last COMPTEL evidence for a 
diffused relic extended ($\sim$30 Kpc) halo. A consequence of the model is the 
presence, at huge distances, of NSJ barionic relic within an extended dark 
matter halo.

\end{document}